# Anomalous magnetic transition in a disordered quasicrystal approximant with heavy-fermion nature

Takayuki Shiino,[1*†] Farid Labib,[2*‡] Yuma Hirano,[1] Kazuhiko Deguchi,[1] Keiichiro Imura,[1]

Takanori Sugimoto,[3,4,5] Satoshi Ohhashi,[2] Nobuhisa Fujita,[2] An-Pang Tsai,[2] and Noriaki K. Sato[1]

[1] Department of Physics, Graduate School of Science, Nagoya University, Nagoya 464-8602, Japan
[2] Institute of Multidisciplinary Research for Advanced Materials (IMRAM), Tohoku University, Sendai 980-8577, Japan
[3] Center for Quantum Information and Quantum Biology, Osaka University, Toyonaka, Osaka 560-8531, Japan
[4] Advanced Science Research Center, Japan Atomic Energy Agency, Tokai, Ibaraki 319-1195, Japan
[5] Computational Materials Science Research Team, Riken Center for Computational Science (R-CCS), Kobe, Hyogo 650-0047, Japan

Quasicrystal approximant $(Ce_xY_{1-x})Cd_6$ ($0 \leq x \leq 1$) forms a network of corner-sharing octahedra. We report that $(Ce_{0.8}Y_{0.2})Cd_6$ exhibits an anomalous magnetic transition which can be classified neither into the conventional static magnetic ordering nor into spin glasses. The anomalous transition is characterized by the coexistence of a static order and a frequency-dependent sharp positive anomaly in the 3rd-harmonic positive susceptibility. Based on the investigation of the reference systems $CeCd_6$ and $(Ce_{0.05}Y_{0.95})Cd_6$, we speculate that the anomalous transition could be induced by disorder in the possible frustrated Ce-network in the presence of the Kondo effect.

Nearly half a century ago, Anderson suggested the possible emergence of a novel quantum state due to geometrical magnetic frustration [1]. Since then, many experimental and theoretical explorations have been done for various frustrated magnets: for example, one-dimensional zigzag chains (the number of nearest-neighbor magnetic sites: $z = 4$) [2], two-dimensional kagome ($z = 4$) [3] and triangular ($z = 6$) [4] lattices, Shastry–Sutherland lattices ($z = 5$) [5, 6], and pyrochlore lattices ($z = 6$) [7, 8].

A Tsai-type 1/1 approximant crystal (AC) of a quasicrystal (QC) could be a new type of frustrated magnet having corner-sharing "octahedral" geometrical frustration network ($z = 8$): see Ref. [9] for the meaning of "1/1", and when we refer to AC in the following, it means 1/1 AC. Those ACs and QCs are intermetallic compounds composed of identical Tsai-type cluster, which is comprised of five concentric polyhedral shells [see Fig. 1(a)]: from the center, tetrahedron, dodecahedron, icosahedron, icosidodecahedron, and rhombic triacontahedron. The clusters array in a periodic arrangement for AC and in a quasi-periodic way for QC. The periodic arrangement of the icosahedral shells in AC is presented in Fig. 1(b). Mysteriously, the AC completes all Platonic solids (with moderate distortion) in its crystal structure: cube and octahedron connect two adjacent dodecahedra and icosahedra, respectively [10, 11]. Twelve rare-earth (RE) atoms bearing magnetic moments reside on the vertices of the icosahedral shell; from another perspective, they form a network of corner-sharing octahedra [11], as shown in Fig. 1(c). These magnetic moments may cause geometrical frustration, which could induce intriguing states and phenomena characterized by the corner-sharing octahedral lattice that has a large $z$ value of $z = 8$. However, no intriguing magnetism has been detected in the RE-based ACs: they only exhibit either spin-glass (SG)-like freezing [12–15] or static long range magnetic ordering [16–21] at low temperatures.

Geometrical frustration in heavy-fermion systems is also an interesting topic; the competition between the geometrical frustration and the Kondo effect may cause an intriguing phenomenon such as metallic spin liquid as observed in distorted kagome ($z = 4$) [22, 23], and pyrochlore ($z = 6$) [24] Kondo-lattice systems. Despite the variety of RE-based ACs discovered and studied so far, there are only a few Ce- and Yb-based ACs that are known to be heavy-fermion systems: $(Ce_xY_{1-x})$–Ag– In [14, 15] and Au–Al–Yb [25]. These systems have been studied with a focus on the quantum critical phenomena, but the possible octahedral frustration effect in the heavy-fermion ACs has not been clarified.

In this Letter, we report the physical properties of $(Ce_{0.8}Y_{0.2})Cd_6$ AC and the reference materials $CeCd_6$ and $(Ce_{0.05}Y_{0.95})Cd_6$. We discover an anomalous magnetic transition/crossover at $T_X \sim 0.55$ K in $(Ce_{0.8}Y_{0.2})Cd_6$; we call it "transition" hereafter although we still do not know whether it is a phase transition or crossover. The characteristic of the anomalous transition is the coexistence of a static order and a frequency dependent positive peak in the 3rd-harmonic ac susceptibility $\chi_3$. To the best of our knowledge, this $\chi_3$ anomaly is distinctive from conventional SGs and any type of magnetic transition and crossover. From the comparison with the reference materials, we suggest that disorder (introduced by the element transition in the geometrically frustrated system with super-heavy-mass quasiparticles.

* Equal contribution
† Current affiliation: Institut de Ciencia de Materials de Barcelona (ICMAB-CSIC), Campus de la UAB, 08193 Bellaterra, Spain.
‡ Current affiliation: Research Institute of Science and Technology, Tokyo University of Science, Tokyo 125-8585, Japan.

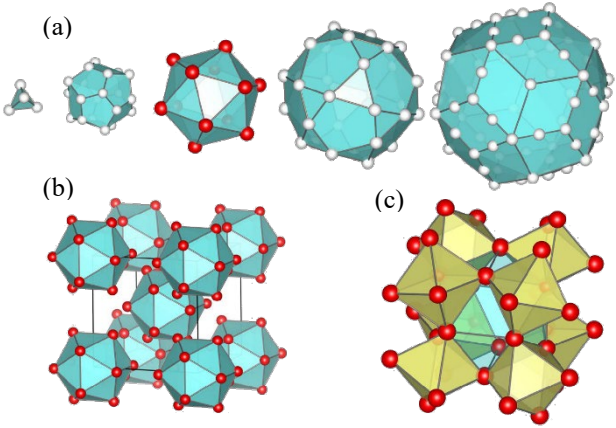

FIG. 1. Schematic illustrations for the crystal structure of Tsai-type $RE$Cd$_6$ 1/1 AC. The white and red spheres represent Cd, and $RE$ atoms, respectively. (a) Tsai-type cluster comprised of five concentric polyhedral shells: tetrahedron, dodecahedron, icosahedron, icosidodecahedron, and rhombic triacontahedron. (b) The arrangement of $RE$ atoms in the bcc structure. (c) Interstitial octahedral lattice units comprised of $RE$ atoms located in between icosahedral clusters.

As-cast nominal $(Ce_xY_{1-x})Cd_6$ alloys, where $x = 1$, 0.8 ,0.05 were prepared from pure elements of Cd (99.999 wt%), Ce (99.99 wt%) and Y (99.999 wt%). Alloying was performed by capsulizing the elements inside stainless-steel tubes employing arc-welding under the Ar atmosphere. The tubes were then sealed within quartz tubes under depressurized Ar gas of 550 Torr. Annealing was performed at 673 K for 100 h in an electric furnace after solutioning at 973 K. The single-phase was confirmed for each sample using the powder x-ray diffraction measurement. Selected-area electron diffraction (SAED) patterns were obtained from transmission electron microscopy (TEM), JEM-2010FEF, with a liquid-nitrogen cooling specimen holder. The dc magnetization $M$ was measured by using a commercial magnetometer (Quantum Design, MPMS) in the temperature range between 1.8 and 300 K and at a magnetic field of $H = 1$ kOe.

The 1st- and 3rd-harmonic ac magnetic susceptibilities, $\chi$ and $\chi_3$ were measured by the conventional mutual inductance method with a commercial $^3$He cryostat (Oxford Instruments, Heliox) and a dilution refrigerator (SAAN, $\mu$-Dilution) below ~2 K using a lock-in amplifier (Stanford Research Systems). A modulation frequency and an amplitude were set in the range of 10.3–2000.3 Hz and 0.1–2.5 Oe, respectively. The ac magnetic susceptibility was calibrated with the dc magnetic susceptibility $\chi = M/H$ with $H = 1$ kOe. We measured the specific heat by using the quasi-adiabatic method with the $^3$He cryostat down to ~0.3 K. We also measured the electrical resistivity using the conventional four-probe method with the $^3$He cryostat and the dilution refrigerator. See Supplementary Material (SM) for further information for the experiments.

The series of $RE$Cd$_6$ ACs can be classified into stoichiometric and off-stoichiometric groups: the off stoichiometric compounds contain additional Cd atoms in the Cd$_8$ cubic interstices [10]. The structure of the stoichiometric $RE$Cd$_6$ (including YCd$_6$) is body-centered cubic (bcc) with the space group of $Im\bar{3}$, while that of off-stoichiometric $RE$Cd$_6$ (including CeCd$_6$, the more accurate stoichiometry of Ce$_6$Cd$_{37}$ [26] thus CeCd$_{6.17}$) is primitive with the space group of $Pn\bar{3}$. The primitive structure of CeCd$_6$ is caused by a perfect crystallographic order of vacant/occupied Cd$_8$ cubic cavities and subsequently orderly oriented Cd$_4$ tetrahedra [10, 26]. Figure 2(a) shows an SAED pattern for CeCd$_6$, confirming the primitive structure. The nominal compound $(Ce_{0.8}Y_{0.2})Cd_6$ of the present interest was revealed to belong to the stoichiometric Tsai-type AC having a bcc structure down to 100 K [see Fig. 2(b)]; see also SM.

Magnetic susceptibility of the present system at high temperatures seems to have a Curie–Weiss component, indicating the localized 4$f$-electron state at high temperatures; note that the non-4$f$-electron contribution to the magnetic susceptibility remains to be clarified (see SM). In the following, we focus on low-temperature physical properties. Figures 3(a)-(d) show the temperature dependence of the physical properties of $(Ce_{0.8}Y_{0.2})Cd_6$ and CeCd$_6$; (a), (c) specific heat $C$; (b), (d) the temperature derivative of electrical resistivity $d\rho/dT$. For both systems, $C$ and $d\rho/dT$ exhibit peaks at 0.5~0.6 K. The peak in $d\rho/dT$ corresponds to the bend in the $\rho$ vs $T$ curve (see also SM), as shown in the insets of Figs. 3(b) and (d), indicating that a static magnetic ordering onsets at $T_X$= 0.5–0.6 K (see below for magnetic susceptibility) as in some other $RE$Cd$_6$ systems

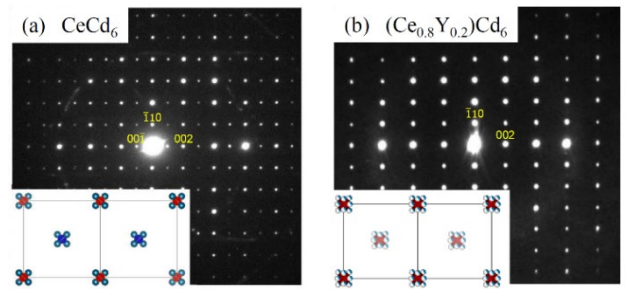

FIG.2. SAED patterns taken from incidences along [110] for (a) CeCd$_6$ measured at room temperature and (b) $(Ce_{0.8}Y_{0.2})Cd_6$ at $T$ = 100 K. No extinction is observed for general $hkl$ reflections of CeCd$_6$, indicating a primitive lattice for CeCd$_6$ at room temperature. The superlattice reflections disappear in the pattern obtained from $(Ce_{0.8}Y_{0.2})Cd_6$ at $T$ = 100 K, satisfying the condition of $h + k + l$ = even; this suggests that $(Ce_{0.8}Y_{0.2})Cd_6$ has a bcc lattice. The insets represent the schematic illustration of the Cd$_4$ tetrahedra: ordered ones for CeCd$_6$ and disordered ones for $(Ce_{0.8}Y_{0.2})Cd_6$.

...

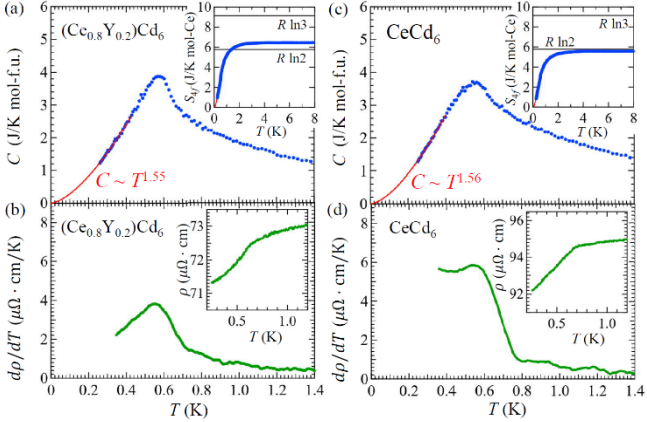

FIG. 3. (a), (b) Temperature dependence of (a) specific heat $C$ and (b) temperature derivative of electrical resistivity $d\rho/dT$ for $(Ce_{0.8}Y_{0.2})Cd_6$. (c), (d) The same for $CeCd_6$. The red solid curves in (a) and (c) indicate fitting results; $C \propto T^{1.55}$ for $(Ce_{0.8}Y_{0.2})Cd_6$ and $C \propto T^{1.56}$ for $CeCd_6$, respectively. The insets in (a) and (c) show the temperature dependence of the entropy of $4f$-electron contribution. The insets in (b) and (d) show the temperature dependence of $\rho$.

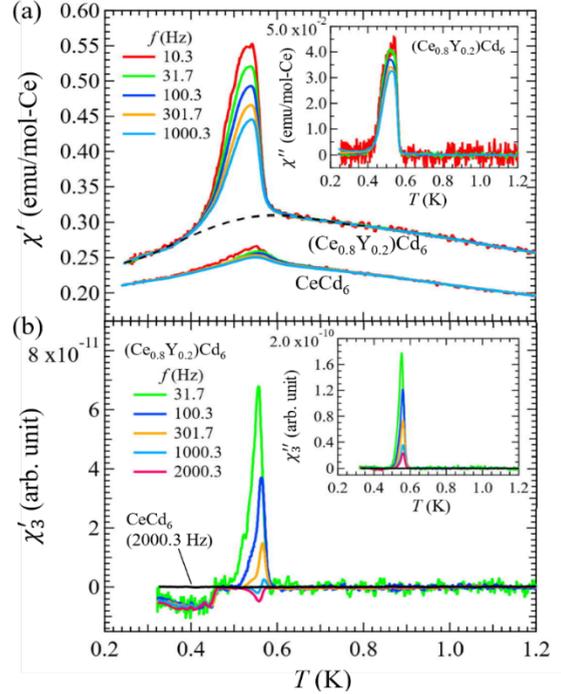

FIG. 4. (a) Temperature dependence of the in-phase part of 1st-harmonic ac susceptibility $\chi'$ for various values of ac frequency. The dashed curve is a guide for the possible frequency-independent component (see the text). (b) The same for the in-phase part of 3rd-harmonic ac susceptibility $\chi_3'$; for $CeCd_6$, the data for 2000.3 Hz is presented. The insets in (a) and (b) show their out-of-phase components. We plotted data of $\chi''$, $\chi_3'$, and $\chi_3''$ by assuming that their values are zero at $T = 1.0$–1.2. In (b) and its inset, we plotted $v_3/fh^3$ [V/(Hz·Oe³)] based on the relation $\chi_3 \propto v_3/fh^3$ [14], where $v_3$ is the induced voltage and $h$ is the amplitude of the applied ac field; note that all data presented in (b) were collected under $h = 2.5$ Oe.

[27]. The broadness of the peaks in $C$ could be due to a characteristic of the present system with possible geometrical frustration. From the fitting of the data to the function of $C = aT^n$ (where $a$ and $n$ are fitting parameters) below ~0.4 K, we find that $n \approx 3/2$ for both $(Ce_{0.8}Y_{0.2})Cd_6$ and $CeCd_6$ [see Figs. 3(a) and (c)]; this indicates the existence of excited states characterized by a quadratic dispersion relation in the statically ordered state. Note that the exponent $n \approx 3/2$ is contrasted from that of SG ($n = 1$) [28], supporting the above identification that the ground state is not a conventional SG.

Using the above power-law temperature dependence of $C$ below the base temperature of the measurement, we calculate the entropy of $4f$-electron contribution ($S_{4f}$) [see the insets of Figs. 3(a) and (c)]. Here the phonon contribution was estimated from the linear slope of $C/T$ vs. $T^2$ plot in the range of $20 \le T^2 \le 60$ K² and subtracted from the measured specific heat ($C$). For both systems, the value of $S_{4f}$ recovers close to $R\ln2$ (where $R$ is the gas constant) well above $T_X$, indicating a Kramers doublet crystal-field ground state; note that the deviation from $R\ln2$ in $S_{4f}$ for $(Ce_{0.8}Y_{0.2})Cd_6$ should be ascribed to a possible error in the extrapolation of the data to $T = 0$ (see SM). Here we remark that the structural difference between $(Ce_{0.8}Y_{0.2})Cd_6$ and $CeCd_6$ (see above) does not reflect in their thermodynamic and transport properties representing the static order.

Figure 4(a) shows the temperature dependences of the 1st-harmonic susceptibility $\chi^0$ (the in-phase component) of $(Ce_{0.8}Y_{0.2})Cd_6$ and $CeCd_6$. The $CeCd_6$ exhibits a broad peak structure with a slight precursor of a frequency-dependent behavior which is almost entirely eliminated by an external magnetic field of 500 Oe (see SM). On the other hand,

$(Ce_{0.8}Y_{0.2})Cd_6$ exhibits an unusual-shaped frequency-dependent peak whose onset has an abrupt increase at ~0.55 K; this peak seems to be superposed on an underlying frequency-independent broad peak as denoted by the guide (dashed curve). The out-of-phase component $\chi''$ also exhibits an anomaly there, as shown in the inset of Fig. 4(a). It seems thus that the $(Ce_{0.8}Y_{0.2})Cd_6$ has two components in $\chi'$; the ergodic (frequency-independent) broad-peak component and the non-ergodic (frequency-dependent) sharp-peak one.

In the case of $(Ce_{0.8}Y_{0.2})Cd_6$, the 3rd-harmonic susceptibility $\chi_3'$ (the in-phase component), which is nearly equal to the 2nd-order non-linear susceptibility [14], exhibits a significant frequency-dependent positive peak near 0.55 K [see Fig. 4(b)]. (Note that the negative anomaly below ~0.46 K should be attributed to the superconductivity of slightly-contained residual cadmium whiskers.) What should be stressed here is that the sign of the peak is positive for $f < 301.7$ Hz. (The sign change of the peak around $f \sim$

...

1000 Hz is unclear; it remains to be clarified.) This result indicates that the anomaly does not arise from conventional ferro-/antiferromagnetic or canonical-SG transitions. What is also important is that the $\chi'_3$ anomaly is absent in $CeCd_6$, indicating the possibility that the anomaly is induced by disorder. Correspondingly, the out-of-phase component $\chi''_3$ of $(Ce_{0.8}Y_{0.2})Cd_6$ exhibits a frequency-dependent peak as shown in the inset of Fig. 4(b). Note that $\chi''_3$ exhibits a larger signal than $\chi'_3$, indicating that this anomaly is accompanied by the energy dissipation. The energy dissipation could be caused by magnetic pinning effects due to the site disorder (and possibly percolation-cluster boundaries) introduced in the Ce-network by the element substitution. The peak of $\chi''_3$ seems to shift toward higher temperature as frequency increases, but we do not observe the Vogel–Fulcher relation which is commonly observed in canonical SGs (see SM).

Figure 5(a) displays the temperature dependence of $C_{4f}/T$ normalized to the mole of Ce for $(Ce_{0.05}Y_{0.95})Cd_6$. Here, $C_{4f}$ was obtained by the subtraction of $C$ of $YCd_6$ from that of $(Ce_{0.05}Y_{0.95})Cd_6$. As seen, the $C_{4f}/T$ increases steadily as the temperature decreases, indicating that the system is a heavy fermion. We observe that the $C_{4f}/T$ curve of $(Ce_{0.05}Y_{0.95})Cd_6$ is fitted with the exact solution of a dilute Kondo model [29] within our measurement range of temperature. Remarkably, the estimated Kondo temperature ($T_K = 0.072$ K) is much smaller than those of conventional heavy-fermion systems ($T_K \sim 1$ K), indicating the super-heavy mass of quasiparticles. The dilute Kondo effect of $(Ce_{0.05}Y_{0.95})Cd_6$ is also confirmed from the magnetic susceptibility, as presented in Fig. 5(b); the measured susceptibility was explained by the Kondo model[30], $1/\chi \propto T + T_K$ for $T \gg T_K$, where $T_K \approx 0.072$ K; note that this $T_K$ value is quantitatively consistent with the $T_K$ evaluated above. From this result, we conclude that $(Ce_{0.8}Y_{0.2})Cd_6$ and $CeCd_6$ are Kondo (heavy-fermion) systems.

Here we discuss the origin of the anomalous transition which emerges in the $(Ce_{0.8}Y_{0.2})Cd_6$ AC. Key points in the present study are summarized as follows. First, the anomalous transition (characterized by the coexistence of the static order and the frequency-dependent $\chi_3$ anomaly) was observed only in $(Ce_{0.8}Y_{0.2})Cd_6$, while $CeCd_6$ exhibits the static order without frequency dependent $\chi_3$ anomaly. Thus, the site disorder in the Ce-network would be important for the anomalous transition. The slight precursor of frequency-dependent behavior observed in $CeCd_6$ [see Fig. 4(a)] could be induced by disorder/defects inevitably existing in the real crystal. Secondly, both $(Ce_{0.8}Y_{0.2})Cd_6$ and $CeCd_6$ form the corner-sharing octahedral network (with defects for the former), which may cause geometrical frustration; both systems exhibit the static order probably because of the large $z$ value. Third, they are both expected to form the Kondo–Yosida singlet at very low temperatures [as indicated by $T_K$ of $(Ce_{0.05}Y_{0.95})Cd_6$], which suppresses the Ruderman–Kittel–Kasuya–Yosida (RKKY) interaction between spins residing on Ce ions. From these results and considerations, it seems reasonable to assume that the anomalous order occurs as a result of the combined effect of the spatial disorder, the geometrical frustration, and the strong correlation among electrons.

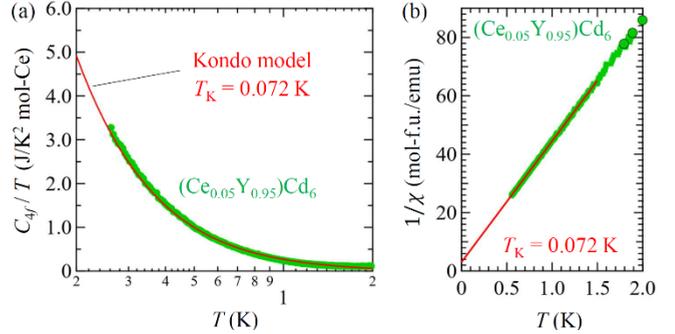

FIG. 5. (a) Temperature dependence of $C_{4f}/T$ of the Cedilute system $(Ce_{0.05}Y_{0.95})Cd_6$. The solid curve is a theoretical curve produced by Rajan's dilute Kondo model [29] with the spin $J = 1/2$ and the characteristic temperature $T_0 = 0.0558$ K. Note that $T_K \approx 1.297 T_0 \approx 0.072$ K [30]. (b) The temperature dependence of inverse magnetic susceptibility of $(Ce_{0.05}Y_{0.95})Cd_6$. The solid curve indicates the linear fitting result extrapolated to the lower temperatures; by assuming the Kondo model $1/\chi \propto T + T_K$ ($T \gg T_K$), we obtain the value of Kondo temperature to be $T_K = 0.072$ K.

Finally, we compare $(Ce_{0.8}Y_{0.2})Cd_6$ with $(Ce_xY_{1-x})$–Ag–In AC [15], which is the only previously reported Ce-based heavy-fermion AC. $(Ce_xY_{1-x})$–Ag–In AC only shows a SG order, and the characteristic temperature $T^*$, which would correspond to $T_K$, is $0.7 \sim T^* \sim 1.4$ K for $0.1 \leq x \leq 1$. Remarkably, $(Ce_xY_{1-x})$–Ag–In has additional inherent chemical disorder (randomness) in specific non-$RE$-element (i.e., Ag and In) sites as well as in the Ce-network, while $(Ce_{0.8}Y_{0.2})Cd_6$ has disorder only in the Ce-network. This fact leads us to conjecture that the disorder in the non-$RE$-sites (i.e., Cd-sites) and/or the Kondo effect with larger $T_K$ could suppress the anomalous transition.

In conclusion, a delicate balance among the disorder, the Kondo effect, and the RKKY interaction in the possible frustrated network of octahedra would be essential for the emergence of the anomalous transition in $(Ce_{0.8}Y_{0.2})Cd_6$. Indeed, the importance of randomness, inhomogeneity, and disorder has been pointed out in many context such as quantum spin liquid [31], heavy fermion [32], and quantum chromodynamics [33]. The geometrical frustration of the octahedral lattice ($z = 8$) must be different from that of the conventional lattices having smaller $z$ values. The heavy-fermion ACs provide a new platform to study the strong correlation effect on the electronic state under the geometrical frustration.

The authors thank Dr. K. Tsuda and Dr. D. Morikawa for his valuable help to conduct the low-temperature TEM experiments and Dr. C. P. Gómez for his advice about the structure of $CeCd_6$. This work was financially supported by

JSPS KAKENHI, Grant Number 15H02111, 19K21847, and 17K18764.